\documentstyle[sprocl]{article}

\def\aj{AJ}
\def\apj{ApJ}
\def\apjl{ApJ}

\def\araa{ARA\&A}

\def\mnras{MNRAS}
\def\pasp{PASP}

\def\sun{\hbox{$\odot$}}

\def\be{\begin{equation}}
\def\ee{\end{equation}}
\def\bea{\begin{eqnarray}}
\def\eea{\end{eqnarray}}

\def\plotfiddle#1#2#3#4#5#6#7{\centering \leavevmode
    \vbox to#2{\rule{0pt}{#2}}
    \includegraphics{#1}}

\def\plotone#1{\centering \leavevmode
    \epsfxsize=\eps@scaling\columnwidth \epsfbox{#1}}

\begin{document}

\title{THE EXTENDED SHAPES OF GALACTIC SATELLITES }

\author{S. R. Majewski, A. D. Forestell, J. C. Ostheimer, C. Palma, M.
  H. Siegel, S. Sohn, K. B. Westfall, R. J. Patterson, P. M. Frinchaboy,
  R. Link}

\address{Dept. of Astronomy, University of Virginia\\
  P.O. Box 3818, Charlottesville, VA 22903-0818, USA\\E-mail:
  srm4n,adf2e,jco9w,cp4v,mhs4p,ssf5b,kbw7d,ricky,pmf8b,rl8z@virginia.edu}

\author{W. E. Kunkel}

\address{Las Campanas Observatory, Carnegie Observatories\\
  Casilla 601, La Serena, Chile \\E-mail: kunkel@jeito.lco.cl}

\author{K. V. Johnston}

\address{Wesleyan University, Department of Astronomy\\
  Middletown, CT 06459-0123, USA\\E-mail: kvj@urania.astro.wesleyan.edu}


\maketitle\abstracts{ We are exploring the extended stellar
  distributions of Galactic satellite galaxies and globular clusters.
  For seven objects studied thus far, the observed profile departs from
  a King function at large $r$, revealing a ``break population'' of
  stars.  In our sample, the relative density of the ``break'' correlates
  to the inferred $M/L$ of these objects.  We discuss opposing
  hypotheses for this trend: (1) Higher $M/L$ objects harbor more
  extended dark matter halos that support secondary, bound, {\it
    stellar} ``halos''.  (2) The extended populations around dwarf
  spheroidals (and some clusters) consist of unbound, extratidal debris
  from their parent objects, which are undergoing various degrees of
  tidal disruption. In this scenario, higher $M/L$ ratios reflect higher
  degrees of virial {\it non-}equilibrium in the parent objects, thus
  invalidating a precept underlying the use of core radial velocities to
  obtain masses.  }

\section{Introduction }

One of the more remarkable aspects of galactic dark matter (DM) is that
the smallest galactic systems, the dwarf spheroidal galaxies (dSph),
appear to have the largest mass-to-light ratios ($M/L$) -- approaching
100 in some Milky Way dSphs.  Even among the dSph galaxies it appears
that intrinsic brightness is anticorrelated with $M/L$, which has
prompted the suggestion that all dSph galaxies have about the same mass,
they just vary in luminosity~\cite{Mat98}.  Globular clusters, the next
smallest stellar systems, seem not to have DM.  It has long been
recognized~\cite{Baade} that a primary difference between dSphs and
globular clusters lies in the concentrations of their radial profiles.
Thus, it is reasonable to consider the {\it structure} of a stellar
system somehow to be connected to the inferred $M/L$'s, although what
the causal relationship may be is unclear: Could the typical, fluffy
profile of dSphs be a response to large DM contents, perhaps in the form
of extended cold DM halos?  Or do more extended {\it stellar} systems
somehow produce {\it the appearance} of large $M/L$'s, even if
artificially?

Deep HST imaging to derive the luminosity function of the Ursa Minor
dSph has shown~\cite{Fel99} no difference with the luminosity function
of the globular cluster M92 down to 0.45 $M_{\sun}$, which establishes
that the apparent {\it stellar} $M/L$ is the same for both low and high
$M/L$ systems.  Thus, it is not surprising that many attempts to
eradicate the ``high $M/L$ problem'' have focused on potential problems
with the methodology for dynamically inferring masses.  Despite general
differences in concentrations, and even shapes, between most globular
clusters and dSphs, it is still commonplace to adopt
King's~\cite{Kin66,RT86} dynamical formalism derived for globular
clusters to describe dSph galaxies.  The assumptions implicit in this
methodology are that the stellar system has an isotropic velocity
distribution and a well-defined, flat core in its light profile (i.e.,
that King profiles fit the light), that $M/L$ is independent of radius,
and that the system is in virial equilibrium.  Concerns have been raised
not only regarding the applicability of each of these assumptions to
dSph systems, but also to whether the dynamics of dSphs are simply more
complex than accounted for by King's model.  For example, several groups
have studied the possibility that the central velocity dispersions,
$\sigma_{v,c}$, are inflated due to superposed orbital motions from
binary stars, but each concludes that binaries alone cannot account for
the large inferred $M/L$'s~\cite{Arm95,Har94,Ols96}.  Improperly
assuming velocity isotropy can also alter inferred $M/L$'s, although
apparently by only factors of two or three, not by factors of 10$^{2-3}$
needed to ``solve'' the dSph $M/L$ ``problem''~\cite{RT86,Mat_araa}.

Perhaps the most controversial assumption has been that of virial
equilibrium.  Hodge \& Michie~\cite{Hod69} long ago proposed that
Galactic tides may act to inflate $\sigma_{v,c}$.  Later
analyses~\cite{Pia95} found that perhaps some, but not all high $M/L$
measures for dSphs can be accounted for by tides, and that, rather than
inflating $\sigma_{v,c}$, tides should produce ordered, shearing-like
motions~\cite{Har94}.  Kuhn \& Miller~\cite{Kuh89} suggested the
possibility that incited resonances between internal stellar orbits and
the bulk Galactic orbit of a stellar system could also inflate
$\sigma_{v,c}$, but Sellwood \& Pryor~\cite{Sel98} counter that such
oscillatory motions are not excited by motion in a logarithmic
potential.  The discovery of potential ``extra-tidal'' stars around some
dSphs~\cite{van78,Sah86,Gou92} raises the specter that tidal disruption
processes, long expected to be acting on some globular clusters, also
affect the supposedly DM-dominated dSphs.  Prodigious disruption, of
course, would be inconsistent with an assumption of equilibrium.  The
Sagittarius (Sgr) galaxy provides an obvious demonstration that Galactic
dwarf satellites can face substantial tidal disruption, and its
derived~\cite{Iba97} $M/L$, in the range of $20-100$, might be
considered a convenient illustration of potential problems with
dynamical masses.  However, Sgr may be a red herring in the dSph DM
argument because its {\it present} morphology is atypical of Galactic
satellites, and it is not obvious that the Sgr progenitor {\it would
  have been} similar to ``normal'' dSphs (though it is likely that Sgr
merely represents an extremely disturbed form of dSph).  On the other
hand, studies of the light profiles of ``normal'' dSphs also reveal signs
of possible tidal effects in the form of radial profile
``breaks''~\cite{Esk88,IH95,Kuh96}, which are predicted as a
manifestation of stripped stars in models of the tidal disruption of
dSph galaxies in a Galactic potential~\cite{JSH99}

The existence of these profile breaks puts interesting twists on the DM
debate.  If the break populations are truly unbound, the degree of
dynamical equilibrium in dSphs is questionable.  Note that Piatek \&
Pryor~\cite{Pia95} show that one pericentric passage of a satellite is
insufficient to perturb a satellite enough to raise the derived $M/L$.
Nor does the presence of unbound stars inflate the measured $\sigma_v$
of their parent systems~\cite{Oh95}.  However, Kroupa~\cite{Kro97} has
commented that earlier disruption models utilized N-body codes too small
to follow the evolution of a satellite to complete dissolution, and that
larger N-body simulations of {\it large} satellites show that late
stages of disruption can produce tidal debris that in places may
converge into 1\% stable remnants that resemble dSphs, but which contain
no DM~\cite{Kle98}.  Such a scenario echos an earlier
suggestion~\cite{Kuh93} that ``large $M/L$ dSphs'' are simply coherent,
unbound groups of stars on similar orbits, the remnants of tidal
disruption.  The notion of dSphs and some globular clusters being
remnant nuggets floating in, and {\it as}, the debris of the disruption
of larger progenitors was first postulated by Kunkel~\cite{Kun79} and
Lynden-Bell~\cite{Lyn82} and more recent analyses discuss the
possibility of at least some ``dynamical families'' of daughter objects
in the outer Galactic halo~\cite{Pal01a}.  But while Mayer et
al.~\cite{Mayer01} confirm that inflated $M/L$ {\it can} occur along
preferred lines of sight to a disrupting system, apparently this might
only account for a small number of the dSphs currently inferred to have
high $M/L$.

However, if the break populations are {\it bound}, it suggests, in the
least, that the structure of dSphs is more complicated than that of
typical globular clusters.  One prosaic explanation for the existence of
bound break populations might be that they are a cloud of still bound
retrograde rotating stars, the product of the dynamical imbalance in
normal evaporative processes to preferentially strip prograde orbiting
stars~\cite{Inn79}; such a model should lend itself to an easily
recognizable observable kinematical signature at the tidal radius,
$r_t$.  Alternatively, the break populations may represent an extended,
secondary ``halo'' component, still embedded in deep dSph DM halos.
According to Burkert~\cite{Bur97}, dSphs need these large DM halos,
since the cut-off radii of King profile fits to dSph are {\it too small}
to be the true $r_t$ if the inferred $M/L$ are correct.

Searching for and following these break populations to extremely large
radii is therefore critical, in order to (1) check on their ubiquity,
particularly among the large $M/L$ systems, (2) look for correlations in
the properties of these break populations as a function of the
characteristics of the core of the system, and (3) see if the break
populations organize themselves, eventually, into the expected tidal
streamers.  The existence of tidal tails and their association to the
break populations is a key discriminant to the various high $M/L$ models
discussed, since ``even modest amounts of dark matter will be very
effective at containing visible stars and halting production of tidal
tails''~\cite{Moo96}.  We have set out to explore the extended profiles
of Galactic satellites (dSphs and clusters), with these, and a number of
other, rationale in mind.

\section{Extended Profiles of Galactic Satellites}

Our survey strategy is intended to go beyond merely obtaining a more
accurate rendering of the light profiles of satellites: We aim to
identify stars widely separated and dispersed from the cores of their
parent systems, and from which we may gather dynamical constraints.  The
technique we employ to work in the low density, ``needle in the
haystack'' regime of the outer profiles of stellar systems is based on
multifilter imaging, including use of the $DDO51$ filter centered on the
gravity-sensitive Mgb triplet and MgH lines at 5150 \AA .  The
Washington $M-T_2$ color provides a temperature index against which to
compare $M-DDO51$ colors, which gauge surface gravity (primarily) and
[Fe/H] (secondarily) in late G through early M stars~\cite{Gei84,M00a}.
Thus, by imaging a stellar system to magnitudes not much fainter than
the red giant branch (RGB), we can discriminate, with a high degree of
confidence and relative ease, between giants associated with that system
and foreground disk dwarf stars.  The latter are a primary contaminating
nuisance and the limiting factor in the effective use of simple
starcounting techniques to trace the extended structures of resolved
stellar systems.  The number of field giant contaminants (i.e., giants
that just happen to be at the same color and magnitude as the RGB of the
parent system) is relatively small (but is subtracted off from the
derived density profiles in Figure 1 below).  Removing virtually all
field star ``noise'' enables us to map these stellar systems to well
past their King limiting radii, which is often adopted as the tidal
radius for a stellar system.

Hunting for individual far-flung stars that can be associated with a
parent stellar system allows one to explore the system to
extraordinarily low effective surface brightnesses (some ten or more
magnitudes below sky brightness).  We have demonstrated this technique
in our study of the Carina~\cite{M00b} and Ursa Minor~\cite{Pal01b}
dSphs.  The reality of our identified ``candidate extratidal stars'' has
been investigated in a variety of ways.  An {\it a posteriori} analysis
of the potential contamination of the isolated RGB sample by photometric
errors yields at most an 18\% expected false detection rate within our
Carina sample~\cite{Pal01c}.  This is much lower than the contamination
rate implied by the analysis of our Carina data by Morrison et
al.~\cite{Mor01}; however, we~\cite{Pal01c} have enumerated a number of
assumptions and other problems (foremost among them, utilization of a
greatly inflated error distribution for our sample) with the Morrison et
al.\ analysis that invalidates their conclusions.  Our estimated {\it
  low} false detection rates are upheld by spectroscopy: (1) Of 50
spectroscopically studied candidate RGB giants in the Carina sample we
find a false detection rate exactly as predicted above (18\%), and we
find nine members beyond the King cut-off radius.  (2) For 155 stars in
the Ursa Minor field with published spectroscopy, we are 100\% accurate
in classifying dwarfs and giants.  The identified extended RGB
population is mimicked by the distribution of Ursa Minor's blue
horizontal branch stars~\cite{Pal01b}.  (3) In similar studies of the
And I and III dSphs, Keck spectroscopy of 54 of our giant candidates
reveals a 100\% identification accuracy~\cite{Guh01}.

We have now made ``associated star'' maps of a number of dwarf galaxy
and globular cluster satellites of the Milky Way and M31 systems.
Figure 1 shows preliminary results on five other satellites with
published $M/L$: the Sculptor, Leo I and Leo II dSph galaxies, and the
globulars NGC 288 and Palomar 13.  For Sculptor, Leo I and NGC 288 we
have employed the techniques discussed above.  For Pal 13, we use a
proper motion membership analysis to identify extratidal
candidates~\cite{Sie01}, while to obtain the density profile of Leo II
we have used the classical method of starcounts~\cite{IH95} on very deep
$UBV$ imaging.

\begin{figure}
\plotfiddle{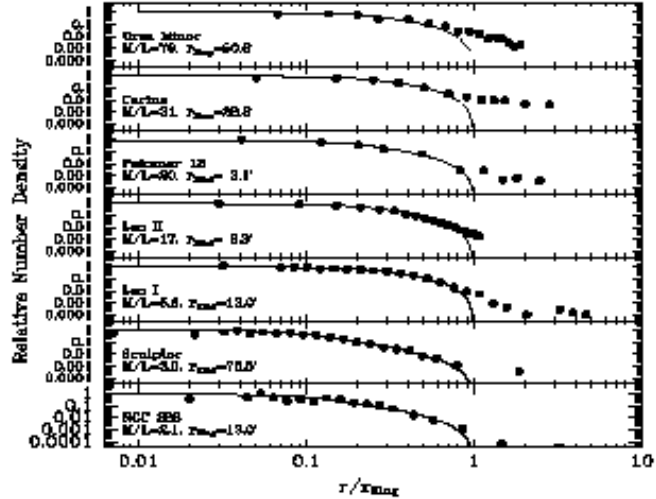}{2.2in}{0}{80.}{80.0}{-230}{-240}
\caption{Radial profiles (normalized to the King limiting radius, 
  often adopted as $r_t$) of the core-normalized density of associated
  star candidates (minus the mean field background) ({\it filled
    circles}) for seven objects, shown in $M/L$ order.  {\it Solid
    lines} show King profiles fit to the centers (all fits are similar
  to, or taken from, previously published fits).  Typical values$^{28}$
  of $M/L$ and $r_t$ are given.  Poissonian error bars are typically of
  order the size of the points.}
\end{figure}

\section{Trend of Departure Density with $M/L$}

All seven systems in Figure 1 show, with varying amplitudes, a break
from their central King profile density distributions that heralds the
onset of a second structural population with a shallower, more or less
power law, fall-off.  We define the ``departure density'' as the
density, normalized to the core, at which the observed profile departs
from the best fit King profile to the central parts.  Among the ensemble
in Figure 1 one sees a trend of higher departure density for higher
$M/L$ objects (Figure 2).  Interpretations at both extremes of the dSph
DM debate can be postulated to account for this apparent trend: (1) dSph
galaxies (and clusters) with DM can support secondary, but bound,
``halo'' populations of stars.  Figure 2 implies that objects with larger
$M/L$ have larger DM halos that can support more substantial stellar
halos.  (2) The extended populations around dSph galaxies (and some
globular clusters) are not bound, but, rather, consist of extratidal
debris from their parent objects, which are undergoing various degrees
of tidal disruption.  In the satellite disruption models of Johnston et
al.~\cite{JSH99}, the departure density reflects the tidal mass loss
rate.  Thus, the apparent correlation suggests that higher inferred
$M/L$'s correspond to higher mass loss rates, and this, in turn,
suggests that the high $M/L$'s are {\it not} a reflection of the DM
contents, but, rather, higher degrees of virial {\it non}-equilibrium in
the parent objects.

\begin{figure}
\plotfiddle{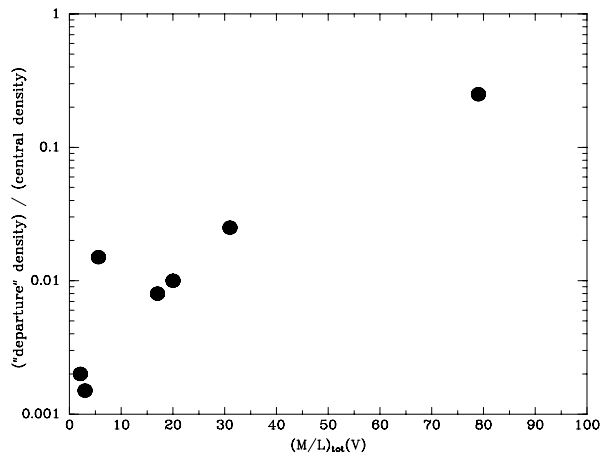}{1.40in}{90}{43}{43}{145}{-50}
\caption{Correlation of departure density to total $M/L$ for Figure 2 
  objects.  The point falling off the otherwise smooth trend represents
  Leo I; our preliminary Leo I analysis may be somewhat compromised by
  the proximity of the star Regulus to Leo I's core.}
\end{figure} 

That Pal 13 partakes in the Figure 2 trend lends some support to the
second scenario.  An $M/L$ of 20 has just been reported for this very
small globular cluster~\cite{Cot01}, while we report elsewhere a number
of reasons why this cluster must be undergoing severe tidal disruption
(e.g., Pal 13 has large blue straggler fraction, a double subgiant
branch, a break in its density profile, and a highly destructive orbit).
In the context of the interpretations for the $M/L$-departure density
trend given above, one must either accept that globular clusters,
including very small ones (with only several thousand $M_{\sun}$ in
stars), can have substantial dark matter contents along with their dSph
counterparts, or that the masses inferred from $\sigma_{v,c}$ are being
substantially inflated due to the effects of tidal disruption.
Fortunately, the extreme interpretations discussed here can be
discriminated by measuring radial velocity trends with radius: DM halos
should produce~\cite{Kle99} declining $\sigma_v$ with $r$ while models
of tidally disrupting systems show~\cite{Kro97} flat or rising
$\sigma_v(r)$.  Our survey, which identifies the very
``needle-in-the-haystack'' targets one needs to do this experiment, has
already allowed us to obtain confirmatory ($\sim15$ km s$^{-1}$
resolution) velocities to significantly larger $r$ (to beyond the King
cut-off radii) than previous surveys (which have been limited to the
high density dSph cores).  Our next challenge is to obtain velocities at
a higher resolution capable of testing the physics.  Models by Kleyna et
al.\ \cite{Kle99} predict that $\le 20$ quality spectra at $\sim 0.75
r_{t}$ are needed to discriminate between models.  These future data
will hopefully bring us closer to understanding the mysteriously large
$M/L$'s of dSph galaxies.
 
We thank Andi Burkert and Ben Moore for useful discussions and support
from the David and Lucile Packard Foundation, The Research Corporation
in the form of a Cottrell Scholar Award, and NSF grant AST-9702521.

\end{document}